\documentclass[letterpaper,10pt]{article}

\date{}

\usepackage[affil-it]{authblk}

\title{Effects of Property Recovery Incentives and Social Interaction on Self-Evacuation Decisions in Natural Disasters: \\ An Agent-Based Modelling Approach}

\author[1]{Made Krisnanda}
\author[2,3]{Raymond Chiong}
\author[1]{Yang Yang}
\author[1]{Kirill Glavatskiy}
\affil[1]{School of Computer and Information Sciences, The University of Newcastle}
\affil[2]{School of Medicine and Public Health, The University of Newcastle}
\affil[3]{School of Science and Technology, The University of New England}

\usepackage[margin=2.5cm]{geometry}          
\usepackage{natbib}                         
\setcitestyle{authoryear,round,aysep={, }}     
\usepackage{booktabs}                        
\usepackage{amsmath,amssymb}                
\usepackage{graphicx}                       
\usepackage{xcolor}                          
\usepackage{float}                           
\usepackage{placeins}                        
\usepackage{url}                             
\usepackage{hyperref}                        
\hypersetup{colorlinks=true, linkcolor=black, citecolor=black, urlcolor=blue}

\begin{document}
\maketitle 

\begin{abstract}
Understanding evacuation decision-making behaviour is one of the key components for designing disaster mitigation policies. This study investigates how communications between household agents in a community influence self-evacuation decisions. We develop an agent-based model that simulates household agents' decisions to evacuate or stay. These agents interact within the framework of evolutionary game theory, effectively competing for limited shared resources, which include property recovery funds and coordination services. We explore four scenarios that model different prioritisations of access to government-provided incentives. We discover that the impact of the incentive diminishes both with increasing funding value and the household agent prioritisation, indicating that there is an optimal level of government support beyond which further increases become impractical. Furthermore, the overall evacuation rate depends on the structure of the underlying social network, showing discontinuous jumps when the prioritisation moves across the node degree. We identify the so-called "community influencers", prioritisation of whom significantly increases the overall evacuation rate. In contrast, prioritising household agents with low connectivity may actually impede collective evacuation. These findings demonstrate the importance of social connectivity between household agents. The results of this study are useful for designing optimal government policies to incentivise and prioritise community evacuation under limited resources.
\end{abstract}

\textbf{keywords:}
agent-based modelling, evolutionary game theory, social network, self-evacuation, disaster management

\section{Introduction}

Natural disasters have clear macroeconomic impacts, with losses reaching \$200 billion per year or more than \$2.3 trillion when cascading and ecosystem costs are taken into account \citep{UNDRR_GAR2025}. Strategic government spending on disaster mitigation measures is fundamental to reducing both human and economic losses in the face of natural hazards. Governments need to invest wisely and ensure they spend on the most crucial yet impactful aspects. Designing a safe and efficient evacuation system that effectively coordinates multiple stakeholders is one of the most challenging and resource-intensive tasks. For instance, governments generally play a leading role in preparing for, responding to and recovering from natural disasters \citep{apo}. Governments provide essential resources, such as information, transportation, policy guidance, and recovery funding, during disaster events and throughout the post-disaster recovery phase. In contrast to the government's broader responsibilities, households have a more immediate and personal objective: protecting their own lives and property. Essentially, households' crucial decision during an emergency is whether to evacuate or stay \citep{Shi2021}.

One of the two key factors influencing household decision-making is that disaster-prone areas typically comprise hundreds to thousands of households that communicate with each other for a period of time before a disaster occurs \citep{Chen2005}. They develop risk perceptions shaped by personal experience and the surrounding social system through these social interactions \citep{Dhar2023}. Householders tend to have varying levels of influence over others based on their roles within the community. For example, prominent community figures, such as local councillors and religious leaders, possess broader communication networks and greater influence on others' decisions \citep{BustillosArdaya2017}. These varying levels of influence and individual interests contribute to complex social dynamics, where interactions among households can collectively shape the community's overall response. 

\citet{Sun2020} examined households' evacuation behaviours during typhoon disasters but did not account for government support through property recovery funds. Governments in developed countries often provide disaster relief programs to support their residents \citep{chester2020}. \citet{Shi2021} considered the interactions between governments and households during wildfires, but hypothesised the rewards for the local government, not the households. However, rewarding the households directly affected by natural disasters is likely to result in higher cooperation during the evacuation process. When disasters strike, the limited emergency response resources and evacuation transportation network capacity constrain the choices available to households. When rebuilding starts, households experiencing property damage can claim recovery funds from the government, subject to budget availability. Providing households with emergency assistance and financial support as rewards can potentially encourage evacuations and contribute to enhanced disaster emergency response plans. However, the effects of such rewards are not yet clear to policymakers. \citet{Rawsthorne2023} emphasised that the ability to map and mobilise social networks is critical to strengthening community action. A deep understanding of community networks and household interactions is crucial for comprehending the historical processes that shape collective actions. \citet{Losee2022} examined the importance of social networks in determining individuals' willingness to respond to environmental threats. Promoting engagement and connection among households increases the disaster preparedness of the whole community. Unfortunately, these studies did not provide detailed explanations of how social networks and their connections can impact the evacuation process. Because physical distance is not a significant factor in network functioning \citep{Flecha2023}, it is appropriate to focus on social relationships of households in the analysis of evacuation dynamics.

The use of modelling and simulation technologies has become a valuable asset in providing scientific data in the conduct of training and exercises. Furthermore, various models, information management tools, and devices are utilised in both exercise environments and to support response operations \citep{Aziz2018}. Mathematical modelling and computer simulation can be used complementarily and jointly to understand the influences of policies. As a natural continuation of this approach, computer simulation can be used to conduct experiments \citep{Mityushev2018}.

Evolutionary game theory (EGT) provides a useful mathematical modelling approach for understanding the evolution of agents' strategies in interactions with potential conflicts \citep{ADAMI20161}. We conceptualise this situation as a social dilemma. Since households tend to make evacuation decisions based on the information available to them \citep{Adedokun2024.2}, we utilised this approach to calculate the benefits and costs of every possible interaction between households. This mechanism mimics human behaviour under emergency situations. 

Agent-based modelling (ABM), as one of the popular computer simulation approaches, has been extensively applied across various climate change-induced disasters, including floods \citep{reviewABMTheRoleandMAS,reviewABMFloodRisk,reviewABMFloodSoTA,reviewABMFloodSimul}, droughts \citep{reviewABMDroughtRisk}, storms \citep{138Hurricanes1,37hurricanes2,72hurricanes3}, and extreme weather events \citep{xtremeWeat}. In this approach, the households interact with each other according to specific rules.  By simulating interactions between households, we can understand how exactly interactions between them impact the overall result. Identifying these mechanisms could benefit the government and policymakers, saving resources and time for the evacuation plan.

Social networks have been utilised to represent interactions in the real world by modelling the relationships and connections between individuals, groups, or entities \citep{Wasserman1994}. These networks are used to map and analyse how information, behaviours, and influences spread through social systems. In our model, nodes represent households, whereas edges represent relationships and interactions between these nodes. We design suitable social network models to reflect various densities and distributions of households in disaster-prone areas. In this study, we utilised a small-world network to represent the community and its interactions as individuals interact in a small-world network through imperfect imitation of behavioural attitudes \citep{torren-peraire2024}. This network model captures specific properties observed in real systems, and understanding their characteristics helps to explain how community structures function. In our experiments, this network represents how people are connected to their neighbours according to the "six degrees of separation" \citep{milgram_small-world_1967}.

\section{Methods}

We develop a model that represents a population residing in a disaster-prone area. Multiple households living in this area regularly interact with one another, both remotely and in person, prior to natural disasters. We use a social network to capture these interactions.  Each node in the network represents an agent, and each edge represents a social interaction between agents. 

We use the Protection Motivation Theory (PMT) framework \citep{Westcott2017} for agent decision considerations. Agents assess the threat (likelihood and severity of a disaster) and the coping appraisal (evacuation and staying costs), to determine which action minimises harm. In addition, the model also considers the effect of limited shared resources (such as road capacity, water, electricity, shelter, and social services) that become critical during evacuation. 

\begin{figure}[H]
    \begin{center}
    \includegraphics[scale=0.4]{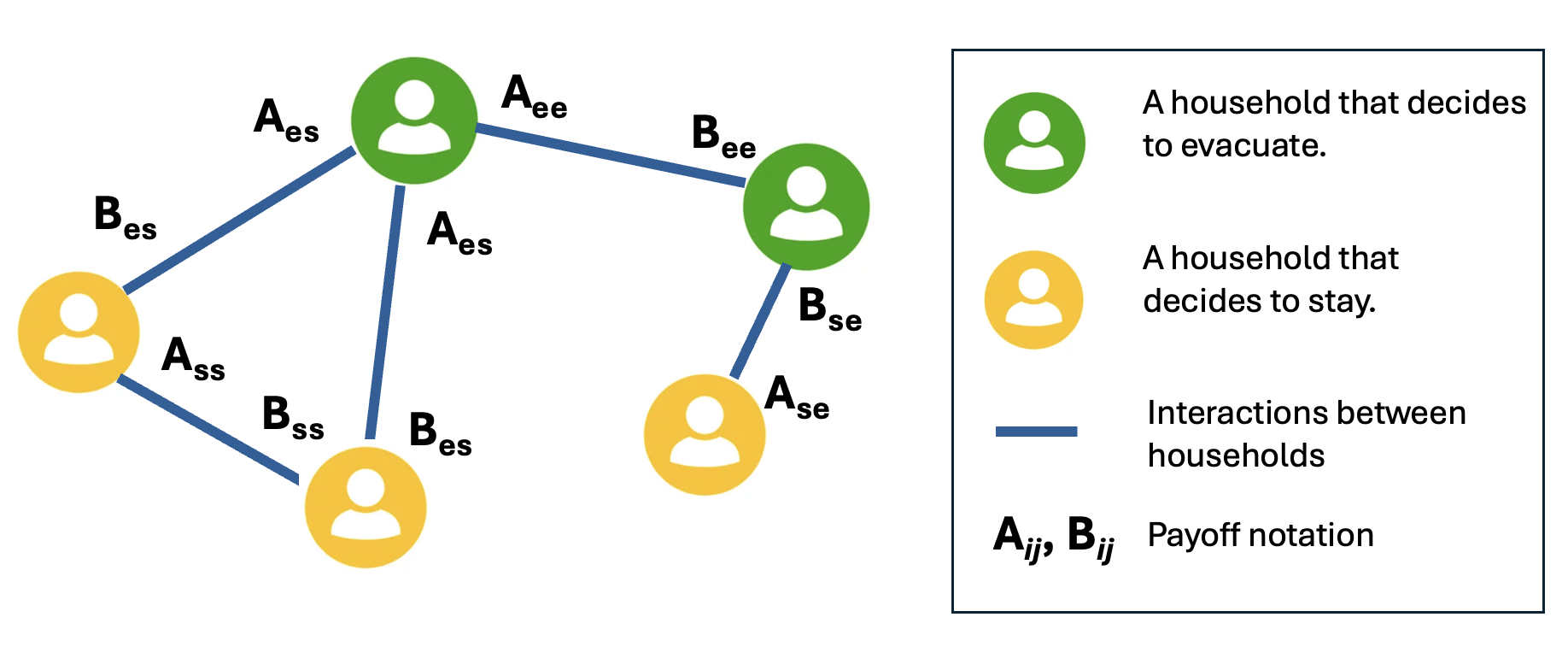}
    \caption{Illustration of interactions between households in the model. The payoff applied depends on the decisions of the two interacting parties.}
    \label{Payoffs}
    \end{center}
\end{figure}

Agents compete for these resources to minimise their perceived losses. For example, if too many agents choose to evacuate, congestion and shelter shortages increase evacuation costs, thereby influencing individual decision-making strategies. We use EGT to model interactions in the population. Using this framework, agents are viewed as players in a game, and their strategies evolve over time depending on their peers success or payoffs. The outcome of the interaction is measured by a payoff, which depends on the agents' decisions. Every pairwise interaction between two agents results in one of eight unique payoff combinations, depending on the perspective of the agent involved. For instance, if some agent A evacuates (subscript "e") and interacts with some other agent B who stays (subscript "s"), then agent A receives a payoff $A_{es}$, and agent B receives $B_{es}$ (see Fig. \ref{Payoffs}). 

We assume that agents treat property as the main factor influencing their evacuation decisions, as financial security is one of the main motivations for people to defend their property \citep{adam2017}.
Each agent owns property and material assets of value $P$. Agents assess the risk of disaster-related losses based on a uniform risk perception value $p$, which represents the expected probability of incurring damage. Before a disaster occurs, agents evaluate the potential damage to their property under both evacuation and stay scenarios. This evaluation assesses how their decisions impact the property's damage.

\subsection{Baseline payoffs}

When two interacting agents both decide to evacuate (or stay), their valuation of the property they own decreases by $pP$, so the expected property value left for each agent is $(1-p) P$. The payoff for each agent in this type of interaction is $A_{ee}^0=B_{ee}^0=A_{ss}^0=B_{ss}^0=(1 - p)P$.

Agents also take into account the costs associated with both evacuating and staying, depending on the decision they make. In particular, if one of the interacting agents is evacuating and the other is staying, the evacuating agent is exposed to an additional cost of $\alpha P$ due to damage to unattended property (e.g., looting, lack of protection), while the remaining agent is exposed to an additional cost of $\beta P$ due to not having help from their neighbours in protecting their home. Additionally, evacuating agents benefit from reduced road congestion and lower shelter crowding, which increases their payoff by $r_E \alpha P$, while staying agents benefit from increased availability of shared resources, which increases their payoff by $r_S \beta P$. As a result, the payoff for evacuating agents is $A_{es}^0=B_{se}^0=(1 - p)P - (1 - r_E)\alpha P$, while the payoff for staying agents is $A_{se}^0=B_{es}^0=(1 - p)P - (1 - r_S)\beta P$. 

Table~\ref{BaselinePayoffCxxx} summarises the resulting baseline payoffs from all possible pairwise interactions between agents in the evacuation and stay groups.

\begin{table}[H]
\setlength{\tabcolsep}{12pt}
\centering
\caption{Baseline Evacuation Payoff} \label{BaselinePayoffCxxx}
\begin{tabular}{ll}
\toprule
\multicolumn{2}{l}{\textbf{Both evacuate} } \\ [5pt]
$A_{ee}^{0}$ = $(1-p)P$ & $B_{ee}^{0}$ = $(1-p)P$ \\
\midrule
\multicolumn{2}{l}{\textbf{A evacuates, B stays}} \\ [5pt]
$A_{es}^{0}$ = $(1-p)P - (1 - r_E)\alpha P$ & $B_{es}^{0}$ = $(1-p)P - (1 - r_S)\beta P$ \\
\midrule
\multicolumn{2}{l}{\textbf{A stays, B evacuates}} \\ [5pt]
$A_{se}^{0}$ = $(1-p)P - (1 - r_S)\beta P$ & $B_{se}^{0}$ = $(1-p)P - (1 - r_E)\alpha P$ \\
\midrule
\multicolumn{2}{l}{\textbf{Both stay}} \\ [5pt]
$A_{ss}^{0}$ = $(1-p)P$ & $B_{ss}^{0}$ = $(1-p)P$ \\
\bottomrule
\end{tabular}
\end{table}
 
\subsection{Government Incentives}

In a typical self-evacuation condition, government support is often available \citep{nswFloodRecovery,redcrossProvidingFinancial, USAgov} , and can be presented in the form of property recovery funds, transportation coordination, shelters, and services. The agents are informed about these incentives in advance and take them into account when making decisions and communicating them during interactions with their neighbours. The government incentives are divided into two categories: financial support and services. The financial incentive is provided in the form of property recovery funds, calculated as a fraction $\theta$ of the agent's property value and aiming to compensate for the property damage. This adds $\theta P$ to the payoffs of all interacting agents who choose to evacuate.

In terms of services, the government provides two types of support: transportation coordination for evacuees and property protection assistance for those who stay. The evacuating agents get an additional benefit from transportation coordination, thus reducing their evacuation cost by a factor of $r_T$. If two interacting agents decide to evacuate, their evacuation costs are reduced by $r_T \alpha P$, while if only one interacting agent decides to evacuate, their evacuation costs are reduced by $r_T(1 - r_E) \alpha P$. This will increase their payoffs to $A_{ee} = B_{ee} = A_{ee}^0 + \theta P - (1 - r_T) \alpha P$ and $A_{es} = B_{se} = A_{es}^0 + \theta P + r_T(1 - r_E) \alpha P$ respectively. The staying agents do not benefit from evacuation services, thus their payoffs remain the same as the baseline payoffs, i.e., $A_{se} = B_{es} = A_{se}^0$. Finally, if both interacting agents decide to stay, they both experience an additional cost $r_D \beta P$ due to governmental resources being allocated to the evacuating group, which decreases their payoff to $A_{ss} = B_{ss} = A_{ss}^0 - r_D \beta P$.

Table~\ref{GovernmentPayoff} presents the complete payoffs for each possible pairwise interaction between agents with additional government incentives. Each equation in the table reflects the corresponding combination of agent decisions (evacuate or stay), incorporating both financial and service-based incentives based on the parameters described above.

\begin{table}[H]
\setlength{\tabcolsep}{12pt}
\centering
\caption{Baseline with Government Incentive Payoff (Blue)} \label{GovernmentPayoff}
\begin{tabular}{ll}
\toprule
\multicolumn{2}{l}{\textbf{Both evacuate}} \\ [5pt]
$A_{ee}$ = $A_{ee}^{0} + \textcolor{blue}{\theta P} - \textcolor{blue}{(1 - r_T) \alpha P}$ & 
$B_{ee}$ = $B_{ee}^{0} + \textcolor{blue}{\theta P} - \textcolor{blue}{(1 - r_T) \alpha P}$ \\ 
\midrule

\multicolumn{2}{l}{\textbf{A evacuates, B stays}} \\ [5pt]
$A_{es}$ = $A_{es}^{0} + \textcolor{blue}{\theta P} + \textcolor{blue}{r_T(1 - r_E) \alpha P}$ & 
$B_{es}=  B_{es}^{0}$ \\ 
\midrule

\multicolumn{2}{l}{\textbf{A stays, B evacuates}} \\ [5pt]
$A_{se}$ = $A_{se}^{0}$ & 
$B_{se}$ = $B_{se}^{0} + \textcolor{blue}{\theta P} + \textcolor{blue}{r_T(1 - r_E)\alpha P}$ \\ 
\midrule

\multicolumn{2}{l}{\textbf{Both stay}} \\ [5pt]
$A_{ss}$ = $A_{ss}^{0} - \textcolor{blue}{r_D \beta  P}$ & 
$B_{ss}$ = $B_{ss}^{0} - \textcolor{blue}{r_D \beta  P}$ \\ 
\bottomrule
\end{tabular}
\end{table}

\section{Experimental setup}

We adopt a small-world network topology to conduct simulation experiments to explore the evolutionary dynamics of evacuation decisions. This network structure is selected because it can realistically reflect a high clustering of households yet allows long-range connections between households that mirror real-world social dynamics \citep{Sallaberry2013}. The network consists of 5,000 nodes representing agents. The degrees of the network range from 2 to 9, see Fig.~\ref{Distribution}.

We run simulations over a limited duration of 3,000 time steps. At each time step, an agent interacts with its connected neighbours and can update its decision. This setup ensures that all agents have an equal number of time steps to revise their decisions. The next time step is processed only after all agents have updated their decisions from the previous one. 

At each time step, agents evaluate their decisions by comparing payoffs with their neighbours, using a probability-based imitation rule. Specifically, agents adopt the decision of a randomly chosen neighbour if that neighbour's payoff is higher. This decision update rule is inspired by \citep{Chica2019}. At each timestep $t$, every agent $i$ compares its total payoff with each neighbour $j$ (where a neighbour is any agent directly connected to $i$). Agent $i$ may adopt neighbor $j$'s previous evacuation decision with a probability $\max\{0, P_{ij}(t)\}$, where 
\begin{equation}
P_{ij}(t) \equiv \frac{w_j - w_i}{\max_k(w_k) - \min_k(w_k)}
\end{equation}
and $w_i$ represents the total payoff agent $i$ receives from all neighbour interactions at $t-1$, while $\max_k$ and $\min_k$ are calculated over the range of total payoffs across all agents at $t-1$. This ensures agents are more likely to imitate neighbours with higher payoffs, allowing evacuation behaviour to evolve based on local interactions. 

All parameters and their complete definitions used in our model are listed in Table \ref{Definition}. The parameter values used in this study are adapted from \citep{Sun2020}.  
\begin{table}[H]
\centering
\caption{Government Incentives: Notations, Definitions, and Values}
\begin{tabular}{|c|p{12cm}|c|}
\hline
\textbf{Notation} & \textbf{Definition} & \textbf{Value} \\ \hline
$p$ & The estimation of risk and cost of life and properties  & 0.5 \\ \hline
$\alpha$ & Cost fraction for evacuating group  & 0.4 \\ \hline
$\beta$ & Cost fraction for staying group   & 0.2 \\ \hline
$r_E$ & Cost reduction for evacuating agent when interacting with the staying agent. & $0.5$ \\ \hline
$r_S$ & Cost reduction for staying agent when interacting with the evacuating agent. & $0.07$ \\ \hline
$r_T$ & Cost reduction from government transportation services & $0.5$ \\ \hline
$r_D$ & Cost increase from government resource limitation. & $0.4$ \\ \hline
\end{tabular}
\label{Definition}
\end{table}

We use the recovery incentive fraction $\theta$ as the control parameter, which ranges from $-10\%$ to $20\%$. This allows us to test how different levels of government support affect evacuation decisions. 

According to Tables 1 and 2, the payoff is proportional to the property value and can be expressed as 
\begin{equation}
\begin{array}{rl}
A_{xy} &= a_{xy}\, P\\
B_{xy} &= b_{xy}\, P\\
\end{array}
\end{equation}
where the coefficients $a$ and $b$ are given in Table \ref{BaselinePayoffCV} and Table \ref{GovernmentPayoffCV}, respectively. From Table \ref{BaselinePayoffCV}, we see that the payoffs are the same if both interacting agents make the same decision. However, the staying coefficient value becomes larger than the evacuating coefficient value if interacting agents make different decisions. Although the government services reduced all payoffs for those who decided to stay, Table \ref{GovernmentPayoffCV} indicates that the agents' propensity to decide to stay is altered by a financial incentive $\theta$, which reflects the amount of funds provided to encourage evacuation. Furthermore, the additional governmental interventions (accounted for by $r_T$ and $r_D$) also influence agents to prefer evacuation. 

\begin{table*}[htbp]
\setlength{\tabcolsep}{12pt}
\centering
\caption{Baseline Evacuation Payoff (Coefficient Values)} \label{BaselinePayoffCV}
\begin{tabular}{ll}
\toprule
\multicolumn{2}{l}{\textbf{Both evacuate}} \\ [5pt]
$a_{ee}^{0} = 0.5$ & $b_{ee}^{0} = 0.5$ \\
\midrule
\multicolumn{2}{l}{\textbf{A evacuates, B stays}} \\ [5pt]
$a_{es}^{0} = 0.3$ & $b_{es}^{0} = 0.656$ \\
\midrule
\multicolumn{2}{l}{\textbf{A stays, B evacuates}} \\ [5pt]
$a_{se}^{0} = 0.656$ & $b_{se}^{0} = 0.3$  \\
\midrule
\multicolumn{2}{l}{\textbf{Both stay}} \\ [5pt]
$a_{ss}^{0} = 0.5$ & $b_{ss}^{0} = 0.5$ \\
\bottomrule
\end{tabular}
\end{table*}

\begin{table*}[htbp]
\setlength{\tabcolsep}{12pt}
\centering
\caption{Baseline with Government Incentive Payoff (Coefficient Values)} \label{GovernmentPayoffCV}
\begin{tabular}{ll}
\toprule
\multicolumn{2}{l}{\textbf{Both evacuate}} \\ [5pt]
$a_{ee} = 0.3 + \theta$ & $b_{ee} = 0.3 + \theta$ \\
\midrule
\multicolumn{2}{l}{\textbf{A evacuates, B stays}} \\ [5pt]
$a_{es} = 0.4 + \theta$ & $b_{es} = 0.47$  \\
\midrule
\multicolumn{2}{l}{\textbf{A stays, B evacuates}} \\ [5pt]
$a_{se} = 0.47$ & $b_{se} = 0.4 + \theta$ \\
\midrule
\multicolumn{2}{l}{\textbf{Both stay}} \\ [5pt]
$a_{ss} = 0.42$ & $b_{ss} = 0.42$ \\
\bottomrule
\end{tabular}
\end{table*}

The final evacuation rates are calculated as the proportion of agents that chose to evacuate over the last 1000 timesteps. This value is further averaged over 5 simulations.   

To examine how the network structure influences evacuation outcomes, we use four scenarios of the households' prioritisation, where the agent's initial decisions are assigned according to their node degree. We assign a fraction of $\gamma$ nodes, which we will refer to as the "priority nodes", with the initial decision to evacuate. The prioritisation levels range from 0\% to 100\%, representing how many agents initially decide to evacuate.
\begin{itemize}
    \item \textbf{Randomised Highest-Degree} In the first scenario, the initial decisions of the remaining $1-\gamma$ nodes are randomised. The priority nodes are $\gamma$ \% nodes with the highest node degree (agents with the most connections).
    \item  \textbf{Fixed Highest-Degree} In the second scenario, the initial decisions of the remaining $1-\gamma$ nodes are assigned to "stay". The priority nodes are $\gamma$ \% nodes with the highest node degree.
    \item \textbf{Randomised Lowest-Degree} Similar to the first scenario, the initial decisions of the remaining $1-\gamma$ nodes in the second scenario are randomised. However, the priority nodes in this scenario are $\gamma$ \% nodes with the lowest node degree (agents with the fewest connections).
    \item \textbf{Fixed Lowest-Degree} In the fourth scenario, the initial decisions of the remaining $1-\gamma$ nodes are also assigned to "stay". The priority nodes are $\gamma$ \% nodes with the lowest degree.
\end{itemize}

\section{Results and Discussion}

This section presents and discusses the simulation results across the four prioritisation scenarios, focusing on how network structure and initial prioritisation levels ($\gamma$) influence evacuation outcomes, including rates, transitions, cascades, and stability.

\subsection{Randomised Highest-Degree}

\subsubsection{Varying prioritisation levels}

\begin{figure*}[htbp]
    \centering

        \includegraphics[width=0.8\textwidth]{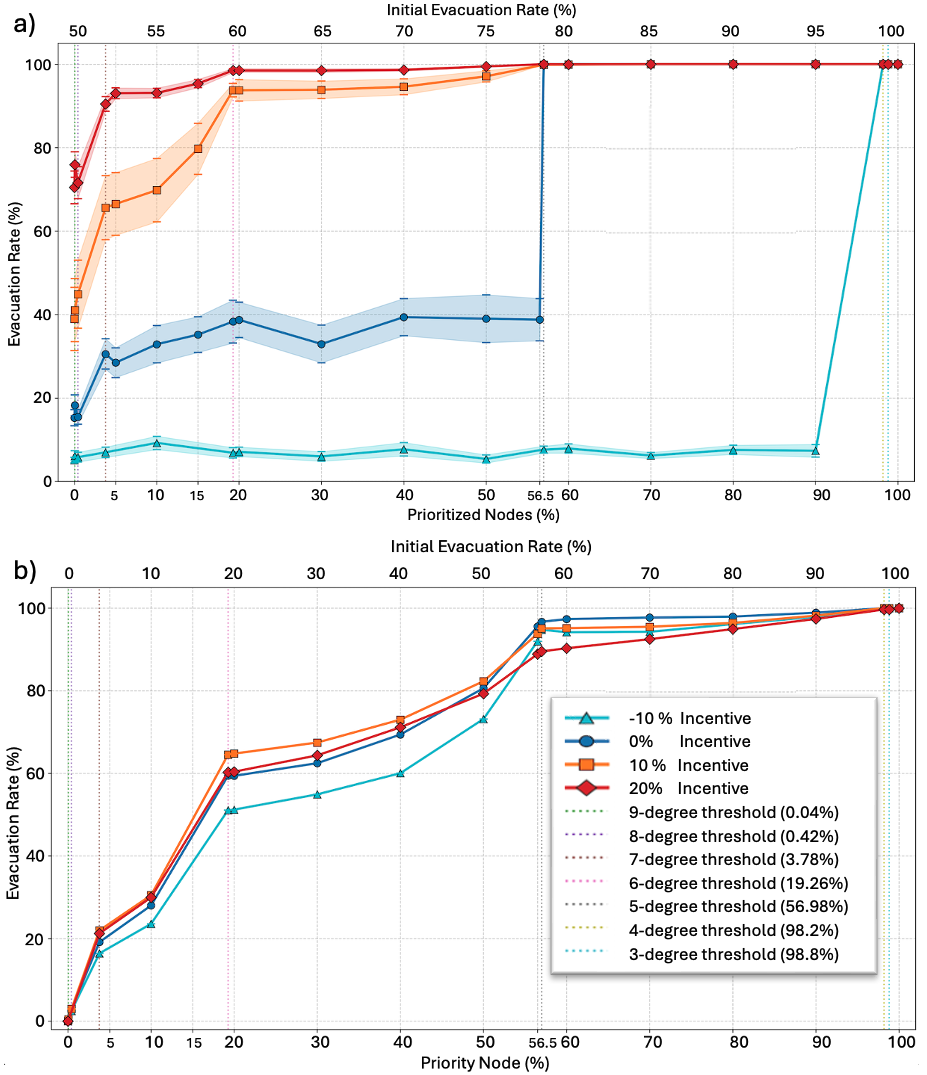} \\

     \caption{Evacuation rates in a. a randomised highest-degree scenario and b. a fixed highest-degree scenario.
    Each point is the average of 5 simulations, and the error band has a width of 2 standard deviations. Vertical lines indicate the fractions of nodes corresponding to the degree change, ranging from degree 9 to degree 3: 0.04\%, 0.42\%, 3.78\%, 19.26\%, 56.98\%, 98.2\%, and 98.8\%.
    These thresholds mark the points at which agents with specific degrees begin to receive prioritised treatment.}
    
    \label{FixedHighest}
\end{figure*}

Figure~\ref{FixedHighest}.a. shows the evacuation rate as a function of prioritisation levels for multiple incentive levels. We observe different increasing trends between degree-based prioritisation thresholds, which separate the groups of agents with the same node degrees.

We first note that at 0\% prioritisation, the evacuation rate has non-zero values for all incentive levels. This is because in the non-prioritised group, there were agents who initially decided to evacuate and hence drove the evacuation decisions of the others. This happened as the agents were aware of the property recovery incentives and could decide to evacuate without additional concern of property damage or looting. 

Next, we see a significant increase in the evacuation rate when the highest degrees (degrees 9, 8, 7) are prioritised. When increasing prioritisation further to 56.98\% (degrees 6 and 5), the agent population involved becomes much larger, leading to a more gradual increase in evacuation. Beyond that, up to 100\% of the prioritised agents are largely low-degree nodes (degrees 4, 3, 2) who, although numerous, do not significantly change the system's state due to their weaker influences on neighbours.

Interestingly, there is an abrupt transition between 56.98\% and 57\% priority in the 0\% incentive case, where the evacuation rate jumped from 38.3\% to 100\% with only a 0.04\% change in priority level. This corresponds to the threshold when agents with degree 4 become involved in prioritisation. At 57\%, agents with degree-5 are fully prioritised, introducing a sudden improvement in connectivity between two previously disconnected clusters (high-degree and low-degree agents). This triggers a cascade of evacuations. 

The same mechanism drives the transition observed at the -10\% level, but at a different prioritisation threshold. There, evacuation remains below 50\% until 98.2\% of the population is prioritised. At this point, 4-degree nodes become fully included, and the evacuation rate jumps to 4,911 agents evacuating simultaneously. 

While the specific thresholds differ, both cases underscore how threshold transitions in the network can lead to instability in evacuation rates.

\subsubsection{Contribution of different degrees}

We next explore the contribution of each degree group to the evacuation rates, as they were progressively included in the prioritisation process. Table \ref{tab:DIFTOP} shows that the role of prioritisation based on the agents' degree is different depending on the incentive level. 
\begin{table}[H]
\centering
\begin{tabular}{c c cccc cccc}
\toprule
\textbf{Degree} & \textbf{Population (\%)} & \multicolumn{4}{c}{\textbf{Rate change}} & \multicolumn{4}{c}{\textbf{Rate change per agent}} \\
\cmidrule(lr){3-6} \cmidrule(lr){7-10}
 & & \textbf{-10\%} & \textbf{0\%} & \textbf{10\%} & \textbf{20\%} & \textbf{-10\%} & \textbf{0\%} & \textbf{10\%} & \textbf{20\%} \\
\midrule
-     & Start  & 5.08   & 15.2   & \textbf{38.9} & \textbf{70.5} & N/A    & N/A    & N/A    & N/A    \\
9     & 0.04   & 1.03   & 3.1    & 2.1           & 5.4           & \textbf{0.51} & \textbf{1.55} & \textbf{1.05} & \textbf{2.70} \\
8     & 0.38   & -0.32  & -2.9   & 3.9           & -4.3          & -0.02  & -0.15  & 0.21   & -0.23  \\
7     & 3.36   & 1.17   & 15.1   & 20.7          & 18.9          & 0.01   & 0.09   & 0.12   & 0.11   \\
6     & 15.48  & -0.12  & 7.8    & 28.2          & 8.0           & 0.01   & 0.01   & 0.03   & 0.01   \\
5     & 37.72  & 0.77   & \textbf{61.7} & 6.1        & 1.5           & 0.00   & 0.03   & 0.00   & 0.00   \\
4     & 41.22  & \textbf{92.39} & 0.0 & 0.0        & 0.0           & 0.04   & 0.00   & 0.00   & 0.00   \\
3     & 0.60   & 0.0    & 0.0    & 0.0           & 0.0           & 0.00   & 0.00   & 0.00   & 0.00   \\
2     & 1.20   & 0.0    & 0.0    & 0.0           & 0.0           & 0.00   & 0.00   & 0.00   & 0.00   \\
-     & End    & 0      & 0      & 0             & 0             & N/A    & N/A    & N/A    & N/A    \\
\bottomrule
\end{tabular}
\caption{Contribution of evacuation rates from different degrees in the randomised highest-degree scenario. The numbers in each column add to 100\%. The largest number in each column is marked in \textbf{bold}.}
\label{tab:DIFTOP}
\end{table}

For negative incentive (-10\%), most of the evacuation decisions (92.39\%) are driven by agents with 4, which is the largest population group (41.22\% of the overall population). Similarly, when there is no incentive, the evacuation is driven by agents with degree 5, which is the second-largest population group (37.72\% of the overall population). This suggests that when the incentive is very low, prioritisation should target the largest population group. In contrast, for high positive incentives (10\% and 20\%), the largest contribution to evacuation (38.9\% and 70.5\%, respectively) is achieved without any specific prioritisation. This suggests that when the incentive is high, there is no need for target prioritisation.

Regardless of the incentive, agents with degrees 2 and 3 contribute nothing to the evacuation decisions. This suggests that agents with very few peer connections play a minimal role in evacuation dynamics and could potentially be excluded from prioritisation strategies. 

When evaluating the contribution of a single agent (as opposed to the group) to the evacuation decisions, the situation is completely different. The agents with the highest degree, so-called "influencers" (degree 9), when prioritised, are able to "convince" evacuating most of the remaining agents. While their overall contribution to evacuation decisions is rather small, due to their small population size, their efficiency (contribution per agent) is the highest. This suggests that they could be considered for the targeted prioritisation when the resources are limited.

Interestingly, agents with the second-highest degree, so-called "ineffective influences" (degree 8), show a negative contribution to the overall evacuation. It is only when the incentive is 10\%, their contribution is positive. However, these negative values do not necessarily mean the 8-degree agents were not contributing to the evacuation rates. Negative changes in degree 8 agents' involvement may have happened due to already high degree 9 agents' previous contributions in the network that overlap with degree 8 agents' influence.

\subsection{Fixed Highest-Degree}

The evacuation rates shown in Fig. \ref{FixedHighest} b) indicate a gradual increase in the evacuation decision that aligns with the priority values. However, we did not find any sudden jumps as those in the first scenario. Although thresholds still have an impact on the increase in evacuation rates, these impacts were significantly smaller in this scenario. This phenomenon can be explained by a complex contagion mechanism, where an agent might need influence from its neighbours before changing its decision. In the first scenario, randomised initial decisions for the non-prioritised group made a small number of evacuation decision seeds that have small and increasing impacts on the evacuation rates. In this scenario, the non-prioritised group initially decided to stay and eliminate the evacuation decision seed until more nodes were prioritised.

Additionally, we found small values of standard deviation, indicating low variation in the results. This has occurred because all simulations began under the exact conditions of evacuation and stay decision separation, resulting in consistent results across simulations.

\subsection{Randomised Lowest-Degree}

\begin{figure}[H]
    \centering
        \includegraphics[width=0.8\textwidth]{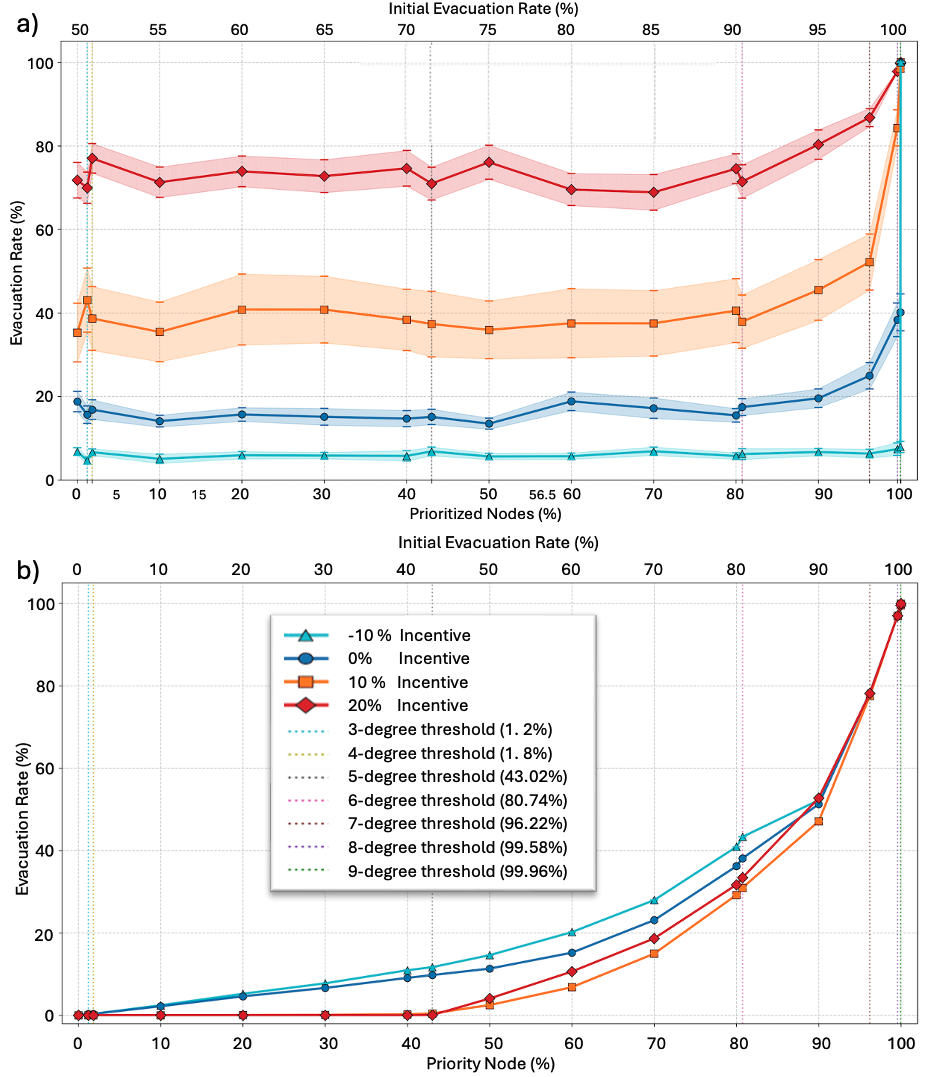} \\
       
         \caption{Evacuation rates in a) a randomised lowest-degree scenario and b) a fixed lowest-degree scenario. In contrast to the highest-degree scenarios, prioritised treatment was initiated from the lowest-degree nodes, ranging from degree 3 to degree 9: 1.2\%, 1.8\%, 43.02\%, 80.74\%, 96.22\%, 99.58\%, and 99.96\%. The results indicate that low-degree nodes make a negligible contribution to the evacuation rates.}
    \label{FixedLowest}
\end{figure}

Unlike the first scenario, the thresholds in this scenario started from 1.2\%. This value represents the population of 2-degree nodes. We then gradually continue with higher degrees up to degree 9. The plot of this scenario is shown in Fig. \ref{FixedLowest} a).

From the figure, we observed no significant change in evacuation rates at all combinations of incentive and priority values. The exception only happened after 80\% of the population was prioritised in all incentive levels. 
The involvement of degree 6 nodes in the prioritisation results in a slight increase in the evacuation rates, especially under 10\% and 20\% incentive values. Moreover, nodes with seven degrees or more experienced a further increase in the evacuation rate, particularly between 90\% and 100\% priority values.
Since the top-degree nodes were prioritised last, this causes a drastic increase in evacuation rates between 90 and 100\% priority values.
In this range, the evacuation rate increases from 5.73\%, 15.5\%, 40.5\%, and 74.5\% to 100\% for incentive values of -10\%, 0\%, 10\%, and 20\%, respectively. 

To understand the contribution of each degree group to evacuation rates in this scenario, we also calculated and grouped the changes in evacuation rates based on the agents' node degree. The details of the results are shown in Table \ref{tab:DIFLOW}. 

\begin{table}[H]
\centering
\begin{tabular}{c c cccc cccc}
\toprule
\textbf{Degree} & \textbf{Population (\%)} & \multicolumn{4}{c}{\textbf{Rate change}} & \multicolumn{4}{c}{\textbf{Rate change per agent}} \\
\cmidrule(lr){3-6} \cmidrule(lr){7-10}
 & & \textbf{-10\%} & \textbf{0\%} & \textbf{10\%} & \textbf{20\%} & \textbf{-10\%} & \textbf{0\%} & \textbf{10\%} & \textbf{20\%} \\
\midrule
-     & Start  & 6.78   & 18.8   & \textbf{35.3} & \textbf{71.8} & N/A     & N/A     & N/A     & N/A     \\
2     & 1.2    & -2.13  & -3.2   & 7.8           & -1.8          & -0.035  & -0.05   & 0.13    & -0.03   \\
3     & 0.6    & 1.99   & 1.2    & -4.4          & 7             & 0.07    & 0.04    & -0.15   & \textbf{0.23} \\
4     & 41.22  & 0.19   & -1.7   & -1.4          & -6            & 0.0     & 0       & 0       & 0       \\
5     & 37.72  & -0.63  & 2.3    & 0.6           & 0.4           & 0.0     & 0       & 0       & 0       \\
6     & 15.48  & 0.11   & 7.5    & 14.3          & 15.2          & 0.0     & 0       & 0       & 0       \\
7     & 3.36   & 1.06   & 13.4   & 32.1          & 11.2          & 0.01    & 0.08    & 0.19    & 0.07    \\
8     & 0.38   & 0.51   & 1.8    & 14.1          & 1.9           & 0.03    & 0.09    & 0.74    & 0.10    \\
9     & 0.04   & \textbf{92.12} & \textbf{59.9} & 1.6     & 0.3           & \textbf{46.06} & \textbf{29.95} & \textbf{0.80} & 0.15 \\
-     & End    & 0      & 0      & 0             & 0             & N/A     & N/A     & N/A     & N/A     \\
\bottomrule
\end{tabular}
\caption{Contribution to evacuation rates from different degrees in the randomised lowest-degree scenario. The largest number in each column is marked in \textbf{bold}.}
\label{tab:DIFLOW}
\end{table}

Based on the results in the table, we found that agents with degree 9 contributed the largest positive change in evacuation rates at both -10\% and 0\% incentive values. Evacuation rates at 10\% and 20\% incentive values were large at time step 0. However, because prioritisation began with the lowest-degree agents, the involvement of degree 2 to degree 5 agents did not significantly change evacuation rates. The higher degree agents (degrees 6 and 7) managed to contribute moderate evacuation rates, but the population could not reach total evacuation until all agents were prioritised.

Focused on the evacuation rate change per agent, degree 9 agents also contributed the most, except at 20\% incentive value. On the other hand, degree 2 agents contributed the least, with negative contributions to the evacuation rates at almost all incentive levels. The only exception was found at 10\% incentive value. This suggests that involving low-degree agents early may delay collective self-evacuation decisions due to poor influence capability.

\subsection{Fixed Lowest-Degree}

The result of the evacuation rate is shown in Fig. \ref{FixedLowest} b). It can be seen that for priorities less than 50\%, the average evacuation rates for all incentive levels are lower in the fixed lowest-degree scenario than in the randomised scenario. This is because both the evacuated and the stayed group decisions were locked at the beginning, especially for the high-degree nodes. Although the lower-degree nodes were prioritised first, they had fewer connections compared to the high-degree nodes and made a smaller impact on the evacuation rates compared to those in scenario 2. 

Moreover, -10\% and 0\% incentive values have higher average evacuation rates compared to other incentive values. This phenomenon happened in almost all priority values, except above 85\%.  The reason is that without any incentive, the stay decision has more payoff compared to the evacuation decision (0.47 vs 0.4), thus making the decision more attractive to the households. On the other hand, with 10\% or 20\% incentive values, the payoffs for the evacuation decision become 0.5 or 0.6 (vs 4.7), making this decision more beneficial for the households.

\section{Conclusions}

This study used ABM to investigate the complexity dynamics of collective evacuation decisions in a small-world social network. The focus is on how the prioritisation strategy and various forms of incentives influence self-evacuation rates. Combined with the EGT, ABM is effective in capturing collective behaviour from individual interactions that are difficult to achieve with traditional modelling approaches. The micro agent interactions within the model reveal non-linearity, thresholds, and complex contagion that are crucial to understanding mass evacuation. 

Based on our experiment, we found that agents don't have equal influences. Prioritising agents with higher connectivity (high-priority nodes) is more effective than prioritising agents with lower connectivity. This result emphasises the importance of social communication in the disaster-prone areas. Identifying and targeting key influencers could trigger an evacuation series, reducing the required resources. 

On the other hand, this study consistently discovered that agents with low social connections (low-degree nodes) contributed minimally to the collective evacuation rates. Prioritising them initially does not significantly boost evacuations and may even impede the self-evacuation process. The findings are essential to policymakers with limited resources, offering a better data-based strategy to maximise public communication and support.

We also found tipping points, such as in 57\% priority values under zero incentive level. Small changes around these points caused sudden jumps in evacuation rates. Moreover, the model revealed a meta-stable state where the population is drifting between two collective decisions: to evacuate or stay. This phenomenon indicates the existence of volatility and sensitivity around thresholds (56.7\% priority value), demonstrating that a small yet effective combination of incentive level and priority value can make a significant difference.

Specifically, the effectiveness of the prioritisation strategy depends on the financial incentive level. Priority values are essential in a limited incentive level. Self-evacuation decisions only spread if the evacuation decision has a higher payoff. On the other hand, prioritisation is not critical if the policymaker has a high level of incentive, and social networks could be a supplement or alternative strategy.

Comparing the random and fixed scenarios shows the importance of initial evacuation decision seeds. In the random scenarios, these seeds have opportunities to grow and accelerate the collective evacuation decisions. On the other hand, fixed scenarios are more resistant to decision changes, requiring more agents or nodes to reach similar results. This highlights the value of early spontaneous evacuees compared to planned actions.

This study provides a new perspective on viewing evacuation dynamics by refuting the simple correlation between the causal mechanisms of peer influence, network structure, and incentives. The findings are crucial to designing more effective and efficient evacuation plans, showing how prioritisation in social networks could be as crucial as monetary values.

However, this study has some limitations that need to be addressed in the future.  For example, the findings are network-specific and validation across diverse topologies (e.g., scale-free, random) is needed. Second, in the real world, social interactions are evolving, and a combination of dynamic networks and an adaptive model is needed to capture this. Last but not least, the agent heterogeneity needs to be expanded to incorporate varying risk perceptions and mobility constraints.

\section*{Acknowledgements}
This work was financially supported by Balai Pembiayaan Pendidikan Tinggi (BPPT), the Indonesia Endowment Fund for Education Agency (LPDP).

\clearpage

\section{Appendix A: Network Structure}

\begin{figure}[H]
    \begin{center}
    \includegraphics[scale=0.4]{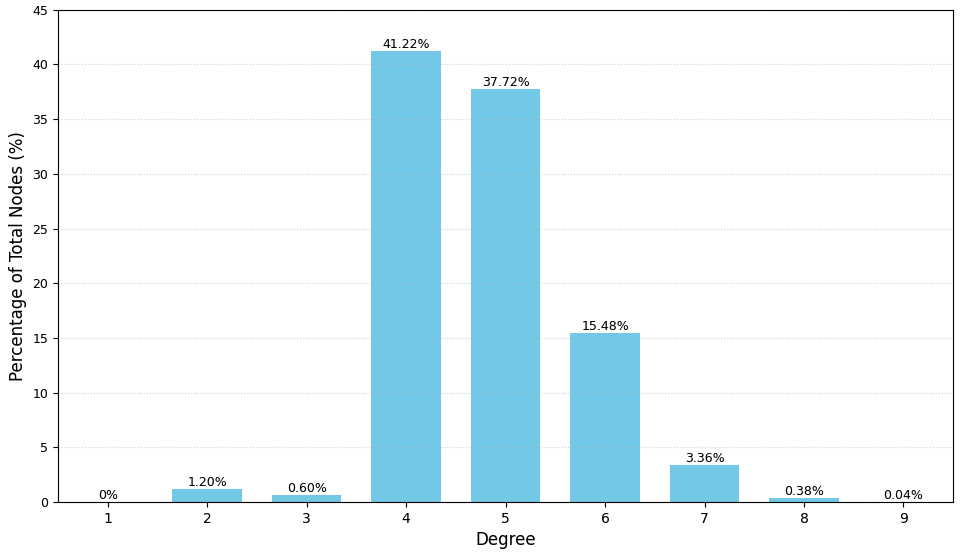}
    \caption{The degree distribution of the small-world network used in the simulations.}
    \label{Distribution}
    \end{center}
\end{figure}

Fig.~\ref{Distribution} shows the degree distribution of the network used in the simulations.

\section{Appendix B: Decision changing dynamics}

Here, we provide further details of the decision-changing dynamics.

\subsection{Decision evolution}

\begin{figure}[H]
    \centering
        \includegraphics[width=0.8\textwidth]{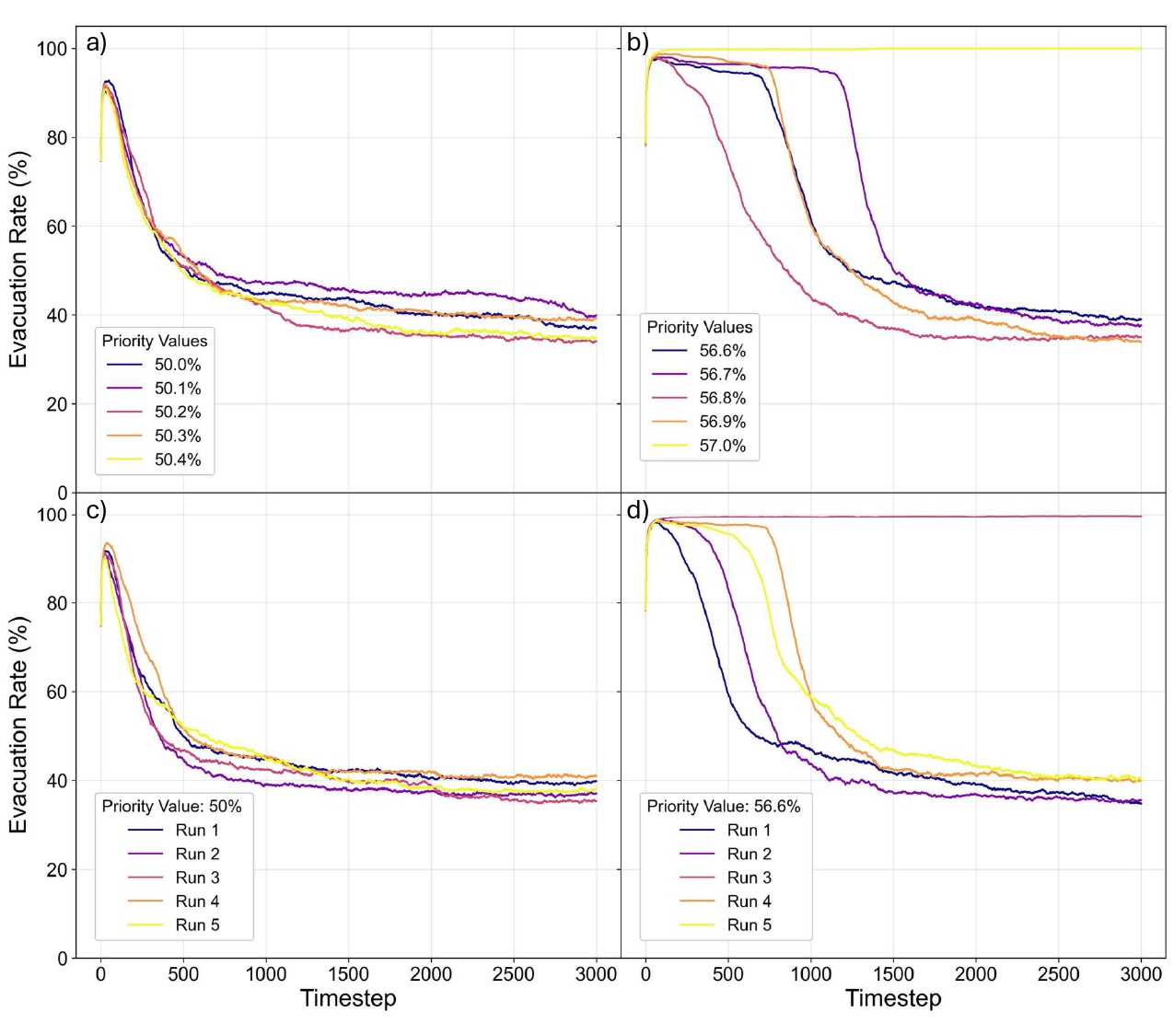} \\
    \caption{Comparison of results around the transition between 50\% and 50.4\% prioritisation(Plot a), and between 56.6\% and 57\% prioritisation(Plot b) in the 0\% incentive values. Plots c) and d) show five curves with the same priority values (50\% and 56.6\%). This figure shows that different results occurred around the thresholds at both different and the same priority levels.}
    
    \label{combined}
\end{figure}

Figure \ref{combined} shows an example of how the evacuation rate evolves in time over 3000 timesteps. The instantaneous evacuation rates were calculated by adding the number of agents that chose to evacuate at each time step and dividing by the total population size, resulting in a percentage over time. 

The figure also shows the evolution of evacuation at a finer resolution between 56.6\% and 57\%, and repeated simulations at 50\% and 56.6\% five times to capture variance under the same conditions. We observe that near the threshold prioritisation, the agents demonstrate a crowd behaviour,
collectively converging on one of two evacuation levels,  either 100\% or approximately 40\%.

This indicates that the evacuation behaviour around the priority threshold can vary sharply, both across different priority values and even at the same priority level.

\subsection{Decision changes}

To investigate the temporal evacuation behaviour across different agent degree levels, we visualise both evacuation rates and decision switching dynamics over time. Although the analysis was performed for all priority values, we were focused on 56\% and 57\% values as representative cases, as they correspond to interesting behaviours identified in earlier experiments. Here, agents were grouped by their node degree (ranging from degree 2 to 9). Within each group, count the number of decision changes across time. This allows us to trace how agents of different connectivity levels respond to evolving conditions, and how volatile their behaviour is under the same incentive and prioritisation setup.

\begin{figure}[H]
    \centering
    
        \includegraphics[width=0.7\textwidth]{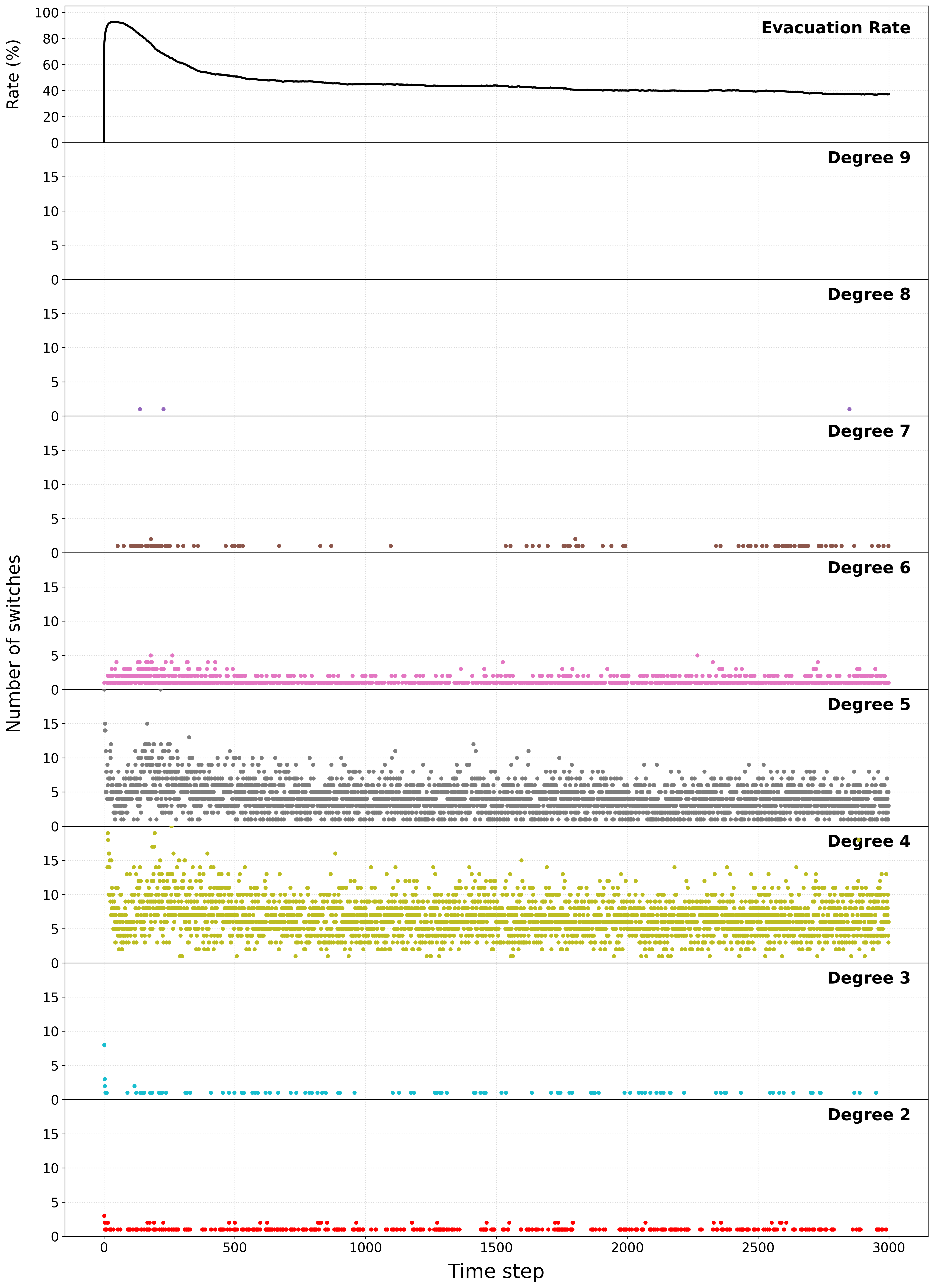} \\
   
    \caption{Evacuation rates and switches result for 0\% incentive and 50\% priority.The top panel displays the overall evacuation rate over time, while the remaining (arranged by descending node degree) show switching frequencies for each degree group.  Each switching plot displays only non-zero switch values using circle markers to highlight discrete decision changes. The x-axis represents simulation time steps, while the y-axis denotes the number of agents switching decisions at each step. }
    \label{ERSW50}
\end{figure}

\begin{figure}[H]
    \centering
   
        \includegraphics[width=0.7\textwidth]{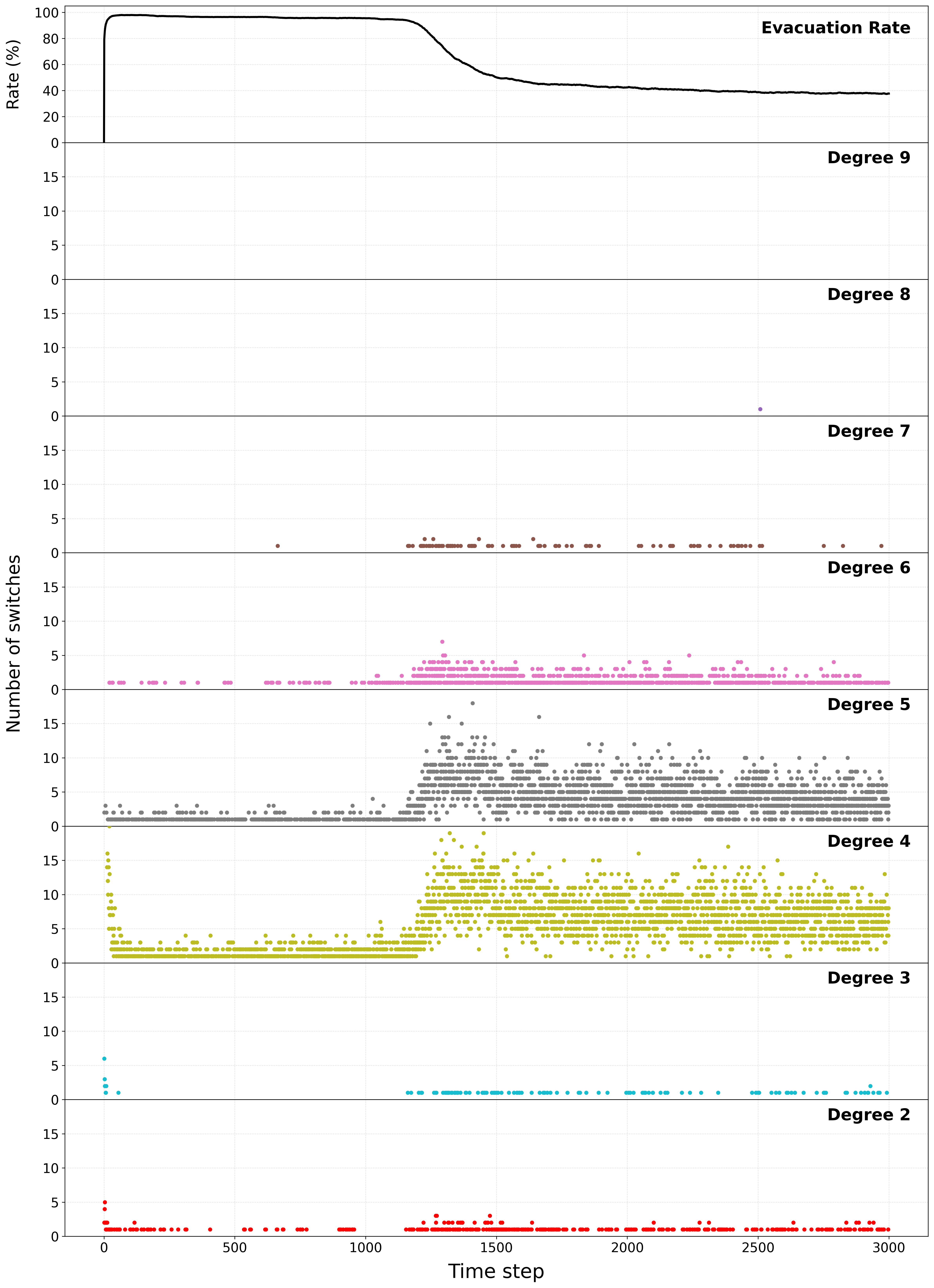} \\
    
    \caption{Evacuation rates and switches result for 0\% incentive and 56.7\% priority.}
    \label{ERSW56}
\end{figure}

\begin{figure}[H]
    \centering
    
        \includegraphics[width=0.7\textwidth]{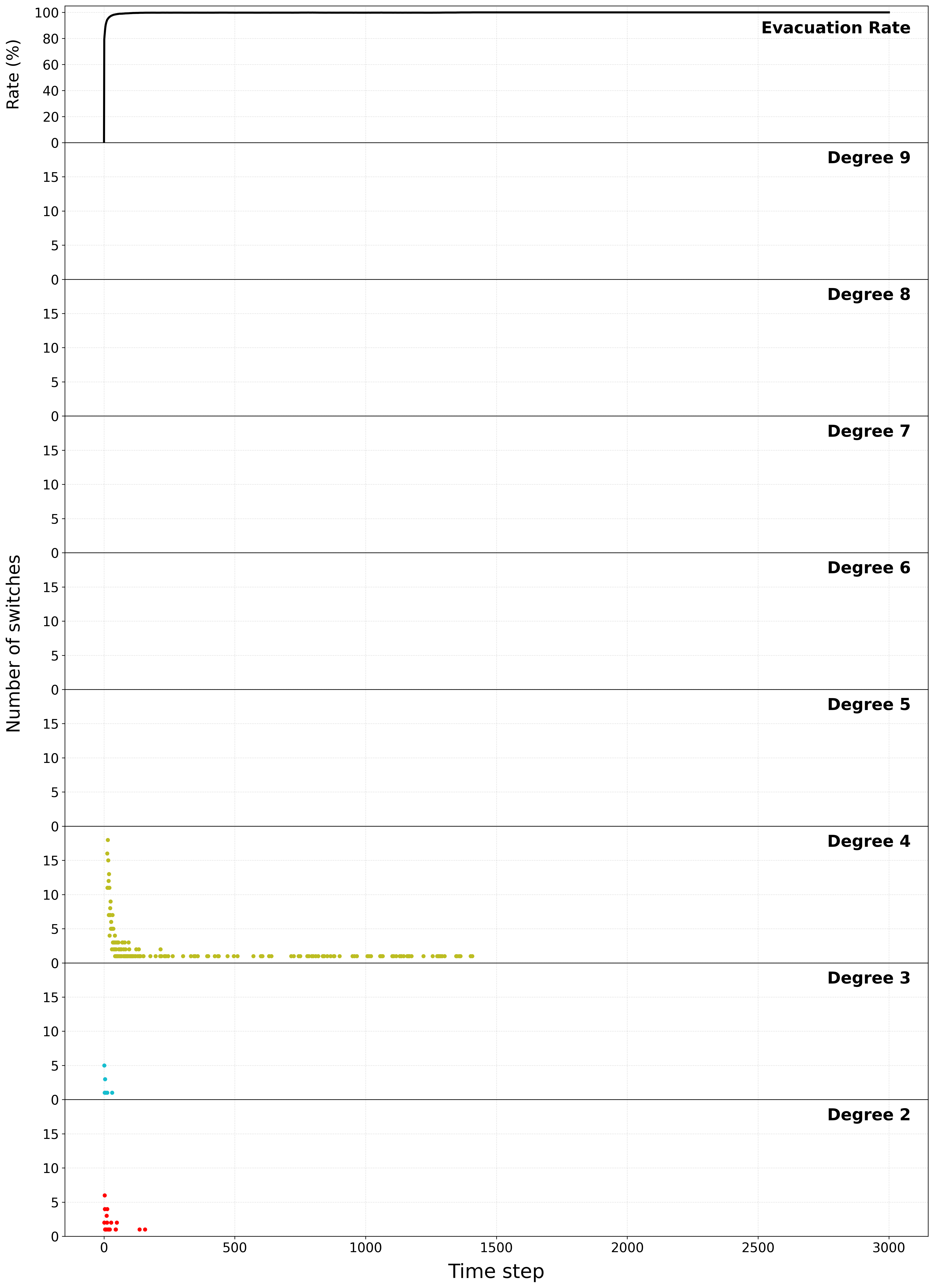} \\
  
    \caption{Evacuation rates and switches result for 0\% incentive and 57\% priority.}
    \label{ERSW57}
\end{figure}

Fig. \ref{ERSW50},\ref{ERSW56}, and \ref{ERSW57} show the decision changes for 50\%, 56.7\% and 57\% priority values respectively and  0\% incentive. 

In the 50\% prioritisation condition, agents changed their decisions a total of 35,158 times. Agents with degree 4, which formed the largest group in the population, made the most changes, accounting for 20,617 of these decision changes. In contrast, agents with degree 8 are the smallest population group and recorded only three decision changes. No agents with degree 9 changed their decisions. When using a 56.7\% prioritisation value, the total number of switches decreased to 26,925. Again, agents with degree 4 made the largest number of decisions. Agents with degree 8 had only one switch recorded. Under the 57\% prioritisation, the total number of decision changes dropped significantly to 1,211. All switching events occurred among agents with degrees less than 5, with degree 4 agents contributing 1,161 of the total. This indicates that decision dynamics are primarily driven by the highest populated group, and higher prioritisation values could stabilise the collective decision by reducing behavioural volatility.

\begin{figure}[H]
    \centering
    
        \includegraphics[width=0.95\textwidth]{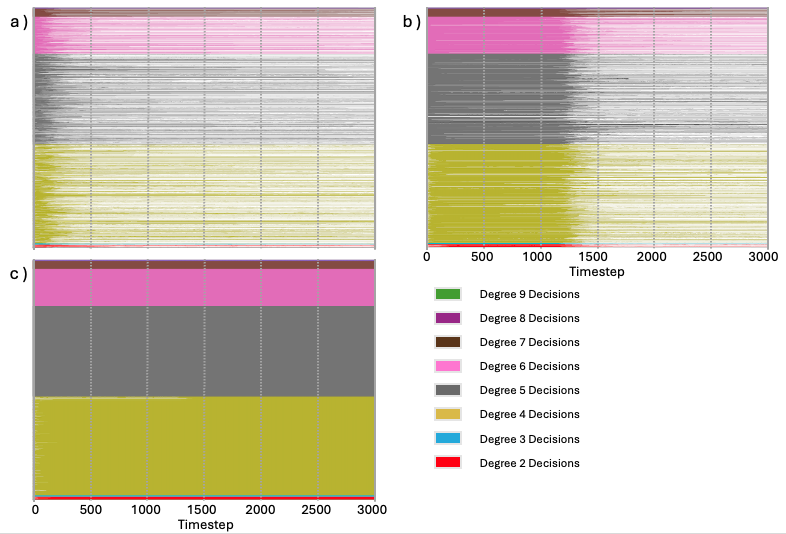} \\
    
    \caption{Heatmap results for 0\% incentive with a.50\%, b. 56.7\% and c. 57\% priority values.}
    \label{HeatmapDecision}
\end{figure}

We also used three heatmaps to observe the visualisation of evacuation and stay decisions across timestep 0 to 3000 for the 50\%, 56.7\% and 57\% priority values (Fig. \ref{HeatmapDecision}). These heatmaps comprise 5,000 agent decisions, clustered by node degree. The x-axis represents the timestep, and the y-axis represents the nodes sorted from the highest to lowest degree(top to bottom). Agents with an evacuate decision are marked with a particular colour depending on the degree, and the stay decision is marked in white. 

At the zero incentive level with 50\% and 56.7\% priority values, both plots show a clear collective transition from evacuate to stay (colored to white). The results will be reserved in the higher incentive level as the payoffs for the evacuation decision increase. In the 50\% priority value, a transition occurred around timesteps 100 to 200, indicating the number of interactions required for the majority of agents to change their decisions. At a 56.7\% priority value, transition occurred around timesteps 1200 to 1300. This indicates that the closer the priority value is to the tipping point (57\%), the population will need more interactions to be stabilised.  In the 57\% priority value, most of the highly influential agents decided to evacuate from the beginning, and all 5-degree nodes were involved in the prioritisation, stabilising the collective decision of the population. In summary, the 50\% priority value results in a collective stay decision, and the 57\% priority value results in a collective evacuation decision. A 56.7\% priority value acts as a meta-stable state, where the population is most volatile and susceptible to flipping between states.

\clearpage 
\bibliographystyle{plainnat}

\typeout{get arXiv to do 4 passes: Label(s) may have changed. Rerun}

\end{document}